\documentclass[fleqn,twoside]{article}
\usepackage{espcrc2}

\usepackage{graphicx}

\begin{document}

\title{High Energy Neutrino Astronomy: The Experimental Road}
\author{Christian Spiering
 \address{DESY Zeuthen, Platanenallee 6, D-15738 Zeuthen, Germany}}

\begin{abstract}
The next ten years promise to be a particularly exciting decade
for high energy neutrino astrophysics.
The frontier of TeV and PeV  energies  is presently
being tackled by large, expandable arrays constructed 
in open water or ice. Detectors tailored to record acoustic, radio, 
fluorescence or air 
shower signatures from neutrino interactions at PeV -- EeV energy 
are being designed in parallel.
During the next decade, the sensitivity to neutrinos from TeV to
EeV energies may improve by 2 to 3 orders of magnitude.
This talk reviews methods, status and prospects of detectors
and sketches a scenario for the experimental progress.
\end{abstract}

\maketitle

\section{Introduction}

\begin{figure}
\centering
\includegraphics[width=6.0cm]{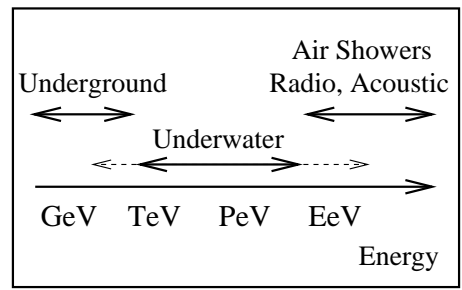}
\vspace{-5mm}
\caption{\label{methodscs}
Energy range of the various detection techniques (see below). Optical 
Cherenkov detectors, although optimized to the TeV-PeV range, are sensitive 
also at lower and higher 
energies, as indicated by the dashed lines.
}
\end{figure}

Whereas MeV neutrino astronomy has been established by 
the observation  of solar neutrinos 
and neutrinos from supernova SN1987,  
neutrinos with energies of GeV to PeV  
which must accompany the production of  high  energy 
cosmic rays still await discovery. 
Detectors underground have turned out to be too small to detect 
the feeble fluxes of  energetic neutrinos from cosmic accelerators. 
The high energy frontier of TeV 
and PeV  energy is currently being tackled by much larger, 
expandable arrays constructed 
in open water or ice. Detectors tailored to record acoustic, radio, 
fluorescence or air 
shower signatures from neutrino interactions at EeV energy 
(= $10^9$ GeV) and above 
are being designed in parallel \cite{Range}. 
Fig.~\ref{methodscs} sketches the energy domains of different techniques.

\section{Physics Goals}

The central goal of high energy neutrino telescopes is to settle 
the origin of high  energy 
cosmic rays \cite{LM,Halzen}. The directional information of these 
charged particles - protons, light and heavy nuclei -  is lost 
due to deflection in cosmic magnetic fields 
(apart from the extreme energies above $10^{10}$ GeV where 
deflection is negligible). Source 
tracing, i.e. {\it astronomy}, is only possible by neutral, stable 
particles like $\gamma$ rays and neutrinos. 
In contrast 
to $\gamma$ rays which may come from pure electron acceleration, 
only neutrinos provide incontrovertible evidence of proton acceleration. 
On top of that, neutrinos do not suffer from absorption by the omnipresent 
infrared or radio background when propagating through space. 
The range of TeV $\gamma$ rays is only about 100 Mpc, 
at PeV even only 10 kpc, i.e. about the radius of our Galaxy. 
Therefore, the topology of the far distant high energy Universe may  
possibly be investigated only with neutrinos.

The physics goals of high energy neutrino telescopes include:

\begin{enumerate}
\item [a)] Search for neutrinos from cosmic acceleration processes in galactic 
sources like micro quasars or supernova remnants (SNR), 
or extragalactic sources like active galactic nuclei (AGN) or 
gamma ray bursts (GRB),
\item [b)] search for ultra-high energy (UHE) neutrinos
from interactions of UHE cosmic rays with the photons of 
cosmic 3K microwave background (the so called GZK neutrinos \cite{GZK}),
from topological defects (TD) or 
from the decay of super-heavy particles,
\item [c)] search for neutrinos from the annihilation of 
Weakly Interacting Massive Particles (WIMPs),
\item [d)] search for magnetic monopoles,
\item [e)] monitoring our Galaxy for MeV neutrinos from supernova bursts.\\
\end{enumerate}

\vspace{-2mm}

Most models related to sources of type 
{\it a)} assume acceleration by shock waves propagating in 
accretion discs around black holes or along the extended jets emitted 
perpendicularly to the disk (bottom-up models). 
Neutrinos are generated in decays of mesons produced by interactions of 
the accelerated charged particles with ambient matter or with photon gas. 

\vspace{-5mm}
\begin{eqnarray*}
p + p (\gamma) \rightarrow   p (n) +  \pi\\
                             &  \searrow  \mu +  \nu
\end{eqnarray*}

The neutrino energy spectrum of many models follows an 
$E_{\nu}^{-2}$ behaviour, 
at least over a certain range of energy. 
Assuming an $E_{\nu}^{-2}$ form and normalizing the neutrino flux 
to the measured 
flux of cosmic rays at highest energies leads to an upper bound of 
$dN/dE_{\nu} \sim 5 \times 10^{-8} E_{\nu}^{-2}$ GeV$^{-1}$ cm$^{-2}$ 
s$^{-1}$ sr$^{-1}$ to the diffuse 
neutrino flux (i.e. the flux integrated over all possible sources) \cite{WB}.
Reasonably weakened assumptions loosen this bound by more than one 
order of magnitude to 
$10^{-6} E_{\nu}^{-2}$ GeV$^{-1}$ cm$^{-2}$ s$^{-1}$ sr$^{-1}$ 
\cite{Berez,MPR}(see also Fig.\ref{diffuse}). 
The so-called top-bottom scenarios of 
type {\it b)} are suggestive for the explanation of highest 
energy cosmic rays. 
In this scenario, high energy particles would be ``born'' 
with high energies, and not accelerated from low to high energies, 
as in the standard bottom-up scenarios. 

We will focus to {\it a)} and {\it b)} in the following and refer to 
\cite{LM,GHS,CS2000,CS2002} 
and references therein for more information on {\it c)} - {\it e)}.

\section{Cherenkov telescopes under water and ice}

Optical underwater/ice neutrino detectors consist of a lattice 
of  photomultipliers (PMs) housed in transparent pressure spheres which are  
spread over a large open volume  in the ocean, in lakes or in ice. 
In most designs the spheres are attached to strings which - in the 
case of water detectors - are moored at the ground and held vertically 
by buoys. The typical spacing along a string is 10-20 meters, 
and between strings 30-100 meters. The spacing is incomparably large 
compared to Super-Kamiokande. This allows to cover large volumes but 
makes the detector practically blind with respect to phenomena below 10 GeV. 

\begin{figure}
\centering
\includegraphics[width=7.3cm]{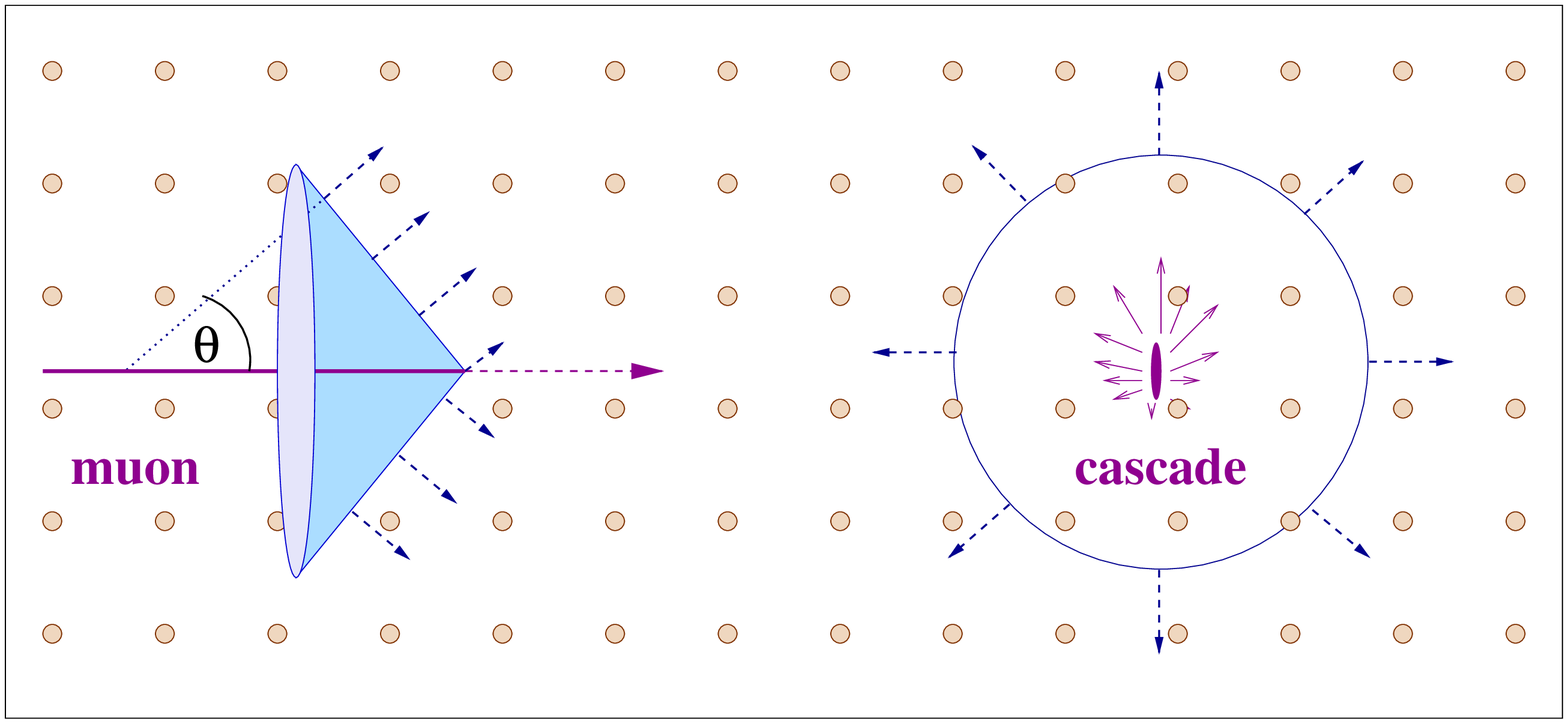}
\vspace{-5mm}
\caption{\label{detmodes} 
Detection of muon tracks (left) and cascades (right)
in underwater detectors.
\vspace{-2mm}
}
\end{figure}

The PMs record arrival time and amplitude of Cherenkov light emitted by 
muons or particle cascades. The accuracy in time is a few nanoseconds. 
Fig.~\ref{detmodes} sketches the two basic detection modes.
 
In the {\it muon} mode, high energy neutrinos are inferred from 
the Cherenkov cone accompanying muons which enter the detector from below. 
Such upward moving muons can have been produced only in interactions 
of muon neutrinos having crossed the earth. 
The effective volume considerably exceeds the actual detector volume 
due to the large range of muons (about 1 km at 
300 GeV and 24 km at 1 PeV). 
Muons which have been 
generated in the earth atmosphere above the detector and 
punch through the water or ice down to the detector, outnumber
neutrino-induced upward moving muons by several
orders of magnitude and have to be removed by 
careful up/down assignment. 
At energies above a few hundred TeV, where the earth is going to become 
opaque even to neutrinos, muons arrive only from directions close to the horizon, 
at EeV energies even only from the upper hemisphere. 
Most of these muons can be distinguished from down going atmospheric 
muons due to their higher energy deposition.

Apart from elongated tracks, {\it cascades} can be detected. 
Their length increases only like the logarithm of the cascade energy. 
With typically 5-10 meters length, and a diameter of the order of 10 cm,  
cascades may be considered as quasi point-like compared to the spacing of 
photomultipliers in Cherenkov telescopes. The effective volume  
for cascade detection is close to the geometrical volume. 
While for present telescopes it therefore is much smaller than that 
for muon detection, for kilometer-scale detectors and not too large 
energies it can reach the same order of magnitude like the latter.

Underwater/ice telescopes are optimized for the detection of muon tracks 
and for energies of a TeV or above, by the following reasons: 

\begin{enumerate}
\item [a)] The flux of neutrinos from cosmic accelerators 
is expected to be harder than that of atmospheric neutrinos above 1 TeV, 
yielding a better signal-to-background ratio at higher energies.
\item [b)] Neutrino cross section and muon range increase with energy.  
The larger the muon range, the larger is the effective detection volume.
\item [c)] The mean angle between muon and neutrino decreases with energy 
like $E^{-0.5}$, with a pointing accuracy of about one degree at 1 TeV.
\item [d)] Mainly due to 
pair production and bremsstrahlung, the energy loss of 
muons increases with energy. Above 1 TeV, this allows to 
estimate the muon energy from the larger light emission along the track.
\end{enumerate}

\begin{figure}
\centering
\includegraphics[width=6.2cm]{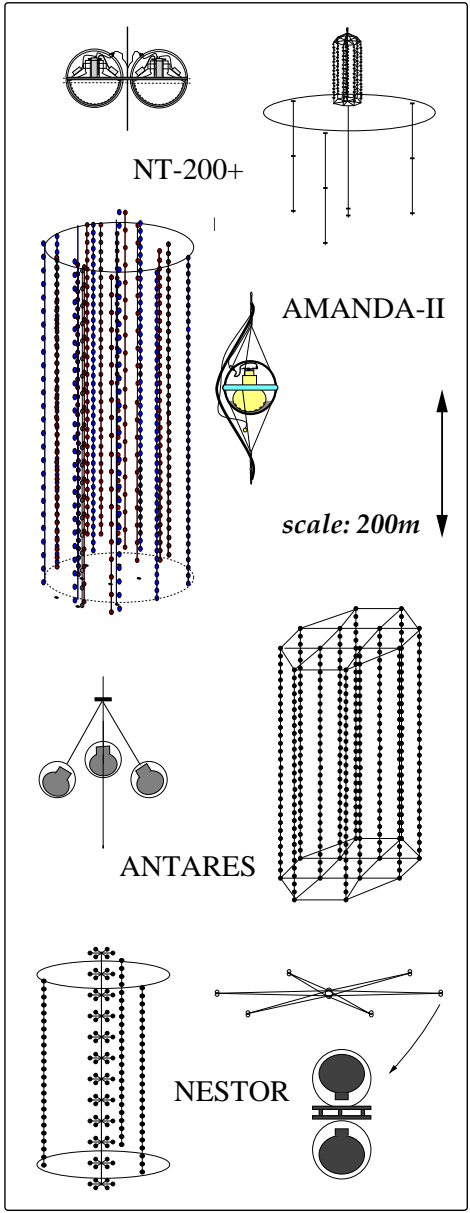}
\vspace{-5mm}
\caption{\label{nutel} 
BAIKAL, AMANDA, ANTARES and NESTOR. Detectors
are shown on the same scale. BAIKAL is shown in
its planned 2004 configuration NT200+,
ANTARES with its 12-string configuration planned
for 2005 and NESTOR with an envisaged ``ring'' made of
old DUMAND modules. For AMANDA, modules shallower than
1.5\,km and deeper than 2.0\,km are omitted in this
figure.
}
\end{figure}

\begin{figure}[h]
\centering
\includegraphics[width=3.1cm]{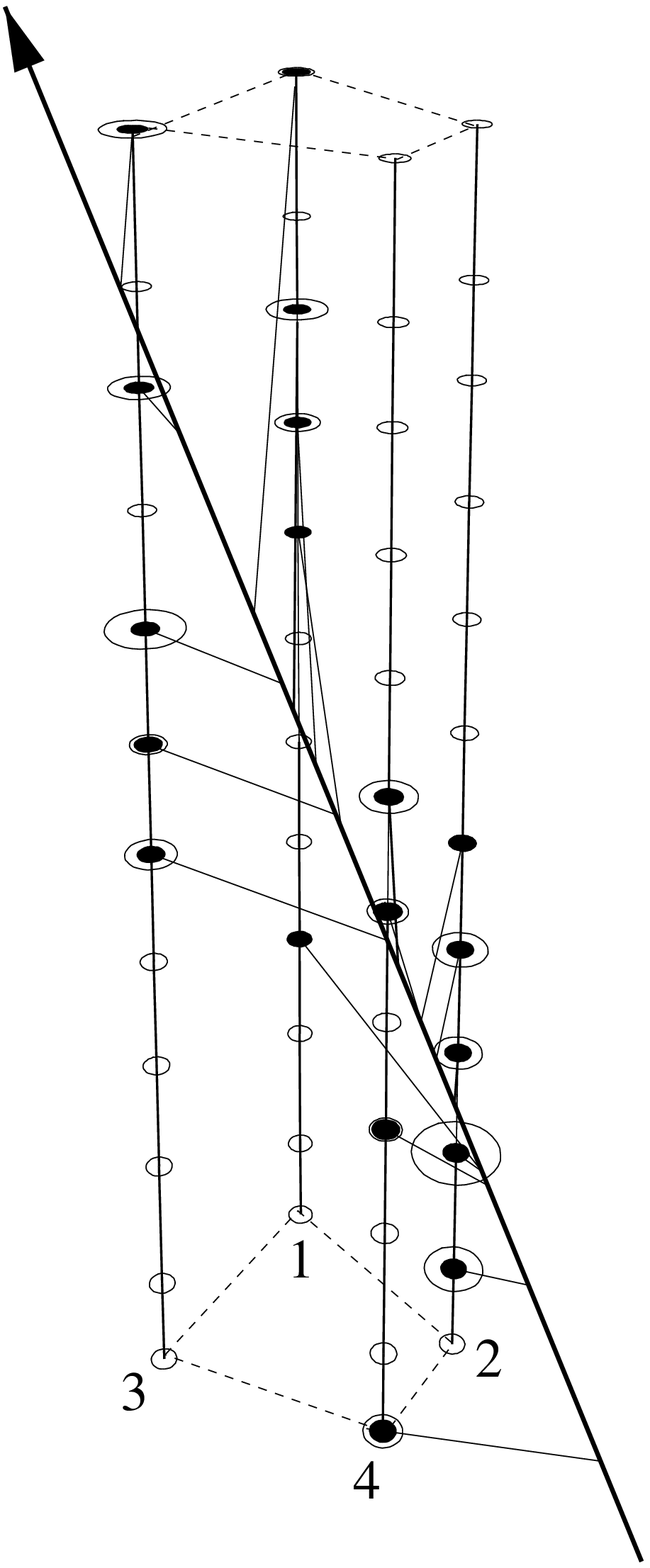}
\hspace{0.7cm}
\includegraphics[width=3.1cm]{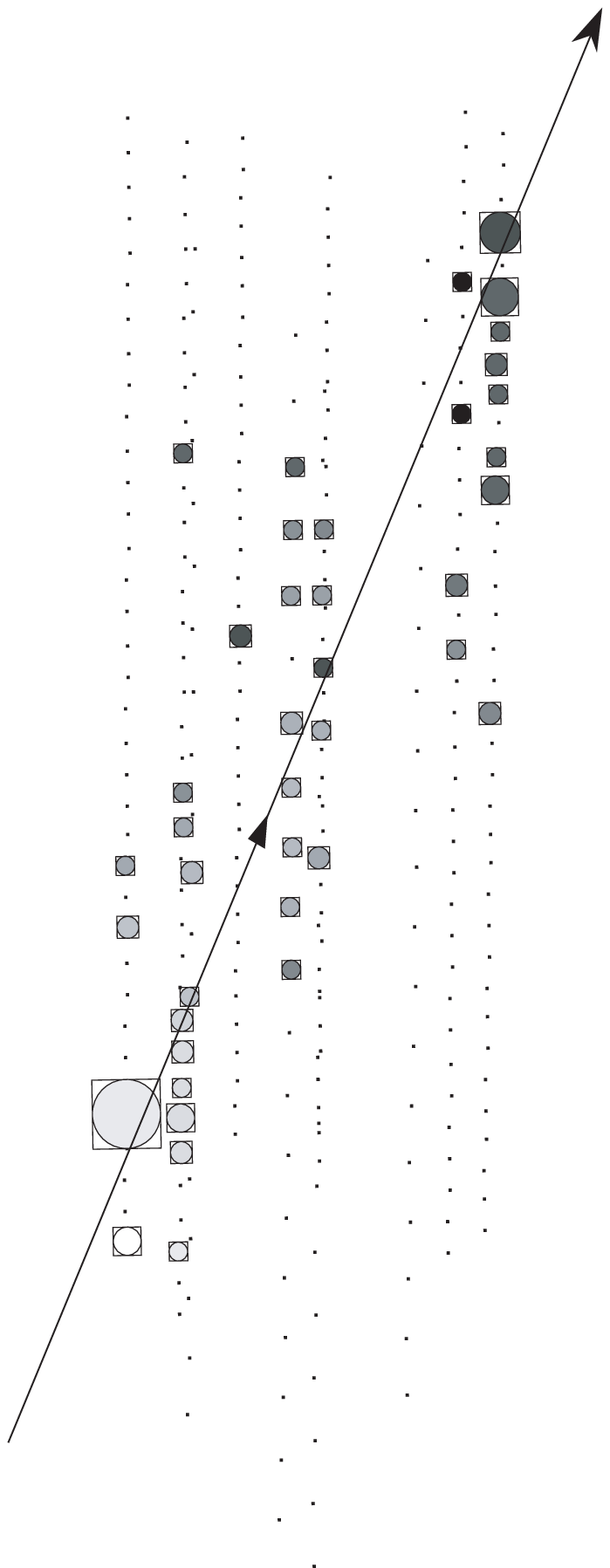}
\vspace{-5mm}
\caption{\label{nuevents} 
Left: one of the first clearly upward moving muons recorded with the 1996 
four-string-stage of the Baikal detector. Small ellipses denote PMs. 
Hit PMs are black, with the size of the disc proportional to the 
recorded amplitude. The arrow line represents the reconstructed muon track,
the thin lines the photon pathes.
Right: Upward muon recorded by the 1997 version of AMANDA. 
Small dots denote the PMs arranged at ten strings.  
Hit PMs are highlighted by boxes,
with the degree of shadowing indicating the time (dark being late),
and the size of the symbols the measured amplitude.
Note the different scales: the height of the Baikal array is 72 
meters, that of AMANDA nearly 500 meters.
}
\end{figure}

The development in this field was stimulated by the {\bf DUMAND} project 
close to Hawaii which was cancelled in 1995. The breakthrough came from 
the other pioneering experiment located at a depth of 1100 m in the 
Siberian Lake Baikal. The {\bf Baikal} collaboration not only 
was the first to deploy three strings (as necessary for full spatial 
reconstruction \cite{BAIKAL1}), but also reported the first 
atmospheric neutrinos 
detected underwater (\cite{BAIKAL}, see Fig.~\ref{nuevents},\,left). 
At present, NT-200 is taking data, an array comprising 192  mushroom 
shaped 15''-PMs at eight strings. 
A moderate upgrade (NT200+) is planned for 2003/04 (see fig.\ref{nutel},
which shows at the top
the small, compact NT-200 array plus three sparsely instrumented 
distant strings forming together NT200+). NT200+ will allow
a significantly improved cascade reconstruction within the
volume framed by the new strings.

\begin{figure}
\centering
\includegraphics[width=7.5cm]{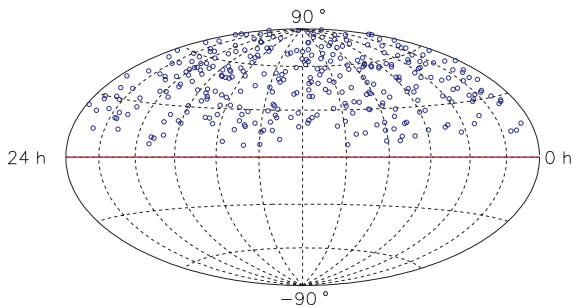}
\vspace{-5mm}
\caption{\label{skymap} 
Sky map of 300 neutrino candidates taken with AMANDA B10 in 1997. 
No indication of 
extraterrestrial point sources on top of  atmospheric neutrinos are found.
}
\end{figure}

With respect to its size, NT-200 has been 
surpassed by the {\bf AMANDA} detector \cite{AMANDA}.  
Rather than water, AMANDA uses the 3 km thick ice layer at the 
geographical South Pole. 
Holes are drilled with hot water, 
and strings with PMs are frozen into the ice. 
With 677 PMs at 19 strings, most at depths between 1500-2000 m, 
the present AMANDA-II array reaches an area of a few 10$^4$ m$^2$ for 
1 TeV muons. Although still far below the square 
kilometer size suggested by most theoretical models, 
AMANDA-II may be the first detector 
with a realistic discovery potential with respect to extraterrestrial high 
energy neutrinos. 
Limits obtained from the analysis of data taken with the 
smaller ten-string detector AMANDA-B10 in 1997 are similar to or below those 
limits which have been obtained by underground detectors over more than a 
decade of data taking. The limit  on the diffuse flux from unresolved sources 
with an assumed $E^{-2}$ spectrum is 
$0.8 \cdot 10^{-6} E_{\nu}^{-2}$ GeV$^{-1}$ cm$^{-2}$ s$^{-1}$ sr$^{-1}$
\cite{diffuse},  
below loosest theoretical bounds \cite{Berez,MPR}, slightly
below the corresponding Baikal limit and nearly an order of
magnitude below limits from underground experiments.  
AMANDA limits on point sources 
on the Northern sky \cite{point} complement the limits obtained 
from detectors on the 
Northern hemisphere for the Southern sky (see Fig.~\ref{point}). 
The sensitivity of AMANDA-B10 has been verified by samples of 
events which are dominated by atmospheric neutrinos \cite{Amanda}.  
Fig.~\ref{nuevents}\,(right)  shows a neutrino event taken 
with AMANDA-B10, Fig.~\ref{skymap} 
the sky map of 
first neutrino candidates taken in 1997.

\begin{figure}
\centering
\includegraphics[width=7.5cm]{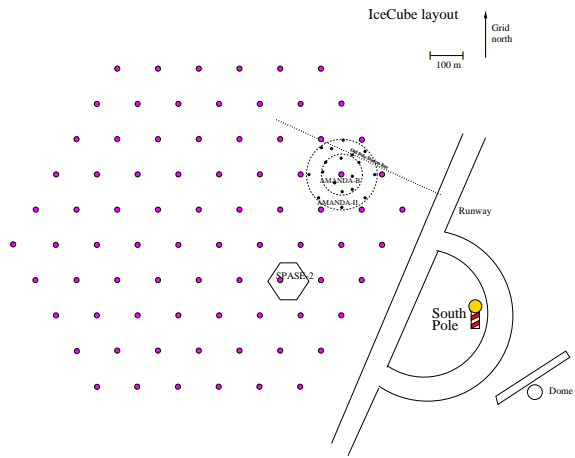}
\vspace{-3mm}
\caption{\label{icecube} 
Top view of the IceCube detector
}
\end{figure}

Based on the experience from AMANDA, a cubic kilometer detector, 
{\bf ICECUBE} \cite{ICECUBE}, is going to be deployed at the South Pole. 
It will consist of 4800 PMs at 80 vertical strings, 
with 125 m inter-string-distances and a 16 m spacing between 
the PMs along a string (see fig.\ref{icecube}). 
The 8-inch AMANDA PMs will be replaced by 
10-inch PMs.  As for the recently upgraded Amanda read-out,
full transient waveforms will be recorded from every PM.

Two projects for large neutrino telescopes are under construction 
in the Mediterranean - {\bf ANTARES} \cite{ANTARES} and  
{\bf NESTOR} \cite{NESTOR} - see fig.\ref{nutel}.  
Both have assessed the relevant physical and optical 
parameters of their sites, developed deployment methods, performed 
a series of operations with a few PMs and layed underwater cables to 
the future locations of the detectors.
ANTARES and NESTOR envision different deployment schemes 
and array designs. The NESTOR group plans to deploy a tower of 
several floors, each 
carrying 12 PMs at 16 m long arms. Later, a ring consisting
of 72 former DUMAND PMs is planned.
The ANTARES detector will consist of 12 strings, 
each equipped with 30 triplets of PMTs. 
This detector will have an area of 
about 2 $\cdot$ 10$^4$ m$^2$ for 
1 TeV  muons - similar to AMANDA-II - and is planned to 
be fully deployed by the end of 2004.  
In addition to these two advanced projects, there is an 
Italian initiative, {\bf NEMO}, 
which finished site investigations at a location  
80 km from Sicily and is now in the 
phase of prototype studies for a cubic kilometer 
detector \cite{NEMO}. At the same time, also 
ANTARES, BAIKAL and NESTOR envisage larger arrays, 
possibly of  cubic kilometer size.

There have been longstanding discussions about the best location for  
a future large neutrino telescope. What concerns geographic location, 
one detector on each hemisphere are necessary for full sky coverage. 
With respect to optical properties, water detectors in oceans seem 
to be favoured: although the absorption length  of  Antarctic ice at 
Amanda depths is nearly twice as long as in oceans 
(and about four times that of Baikal), ice is characterized by strong 
light scattering, and its optical parameters vary with depth. 
Light scattering leads  to a considerable delay of Cherenkov photons. 
On the other hand ice does not suffer from the high potassium content 
of ocean water or from bioluminescence. These external light sources result 
in counting rates ranging from several tens of kHz to a few hundred kHz, 
compared  to less than 500 Hz pure PM dark count rate in ice.  
Depth arguments favour oceans. The seabed at the NESTOR site
is deepest (4 km), closely followed by NEMO 
(3.5 km). With only 2.5 km, the ANTARES ground is 
at about the same depth as the lowest AMANDA modules.
Note, however, that the depth argument lost some of its 
initial  strength after BAIKAL and AMANDA had developed reconstruction 
methods which effectively reject even the high background at shallow depths. 
Actually, the main advantage of great depths is the possibility to
look higher above horizon, i.e. an increased angular
acceptance. For water, a detector at greater depth suffers less from
sedimentation of biomatter and from 
increased noise rates due to bio-luminescence.
What counts most, at the end, are basic technical questions like deployment, 
or the reliability of the single components as well as of the whole system. 
Systems with a non-hierarchical structure like AMANDA (where each PM has 
its own  2 km cable to surface) will  suffer less from single point 
failures than water detectors do.  In the case of water,  longer distances 
between detector and shore station have to be bridged. 
Consequently, not every PM can get its own cable to shore, 
resulting in a hierarchical system architecture. 
This drawback of water detectors may be balanced by the fact that they 
allow retrieval and replacements of failed components, as the BAIKAL 
group has demonstrated over many years.

Most likely, the present efforts will converge to two cubic 
kilometer detectors for very high energy neutrinos, 
ICECUBE at the South Pole and one in the Mediterranean.

\section{Acoustic detection}

\begin{figure}
\centering
\includegraphics[width=4.5cm]{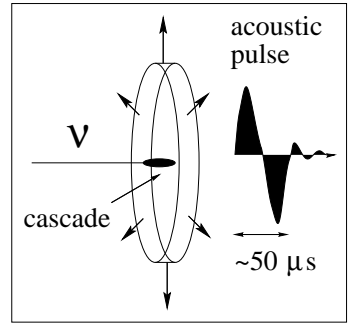}
\vspace{-5mm}
\caption{\label{acoustic} 
Acoustic emission of a particle cascade
}
\end{figure}

Acoustic particle detection was proposed first in the 
fifties \cite{Acoustics1} and experimentally proven 
two decades later \cite{Sulak}. 
A high energy particle 
cascade deposits energy into the medium via ionization losses, 
which is immediately converted into heat. 
The effect is a  fast expansion, generating a bipolar acoustic pulse
with a width of a few ten microseconds in water or ice 
(see fig.\ref{acoustic}). 
Transverse to the pencil-like cascade (diameter about 10 cm)  
the radiation 
propagates within a disk of about 10 m 
thickness (the length of the cascade) into the medium. The 
signal power peaks 
at 20 kHz where the attenuation length of sea water is a few kilometers, 
compared to a few tens of meters for light. Given a large initial signal, 
huge detection volumes can be achieved. Provided efficient noise rejection, 
acoustic detection might be competitive with optical  
detection at multi-PeV energies \cite{Acoustics,Price}.

Present initiatives (see \cite{CS2002}) envisage combinations of 
acoustic arrays with  optical 
Cherenkov detectors (NESTOR, ANTARES, ICECUBE) or the use of 
existing sonar arrays for submarine detection close to Kamchatka and in 
the Black Sea \cite{Capone}. Most advanced is {\bf AUTEC},
a project using a very large hydrophone array of the US Navy, 
close to the Bahamas \cite{AUTEC}.  
The existing array of 52 hydrophones spans an 
area of 250 km$^2$ and has good sensitivity 
between 1-50 kHz. It is expected to trigger on events above 100 EeV 
with a tolerable  false alarm rate.

\section{Radio detection}

Electromagnetic showers generated by high energy electron neutrino 
interactions emit coherent  Cherenkov radiation.  Radio
Cherenkov emission was predicted in 1962 \cite{Radio1}
and confirmed by recent
measurements at SLAC and ANL \cite{Radio2}.
Electrons are swept into 
the developing shower, which acquires a negative net charge from the added 
shell electrons . This charge propagates like a relativistic pancake of 1 cm 
thickness and 10 cm diameter. Each particle emits Cherenkov radiation, 
with the total signal 
being the resultant of the overlapping Cherenkov cones. 
For wavelengths larger than the cascade diameter,  
coherence is observed and the signal rises proportional to $E^2$, 
making the method attractive for high energy cascades. The bipolar radio 
pulse has a width of 1-2 ns.  In ice as well as in salt domes, 
attenuation lengths of several kilometers can be obtained, depending 
on the frequency band, the temperature of the ice, and the salt quality. 
Thus, for energies above a few tens of PeV, radio detection in ice 
or salt might 
be competitive or superior to optical detection \cite{Price}.

\vspace{-2mm}

\begin{figure}[h]
\centering
\includegraphics[width=5.3cm]{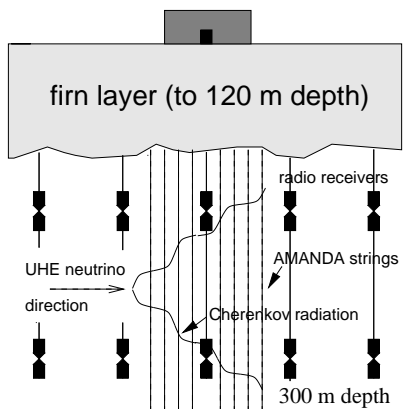}
\vspace{-5mm}
\caption{\label{rice} 
The RICE detector at South Pole
}
\vspace{-0.5cm}
\end{figure}

A prototype Cherenkov radio detector called {\bf RICE} is operating at the 
geographical  South Pole
\cite{RICE}. Twenty receivers and emitters are buried at depths 
between 120 and 300 m (fig.\ref{rice}). 
From the non-observation of very large pulses, a limit of about 
$10^{-4}  E_{\nu}^{-2}$ GeV$^{-1}$ cm$^{-2}$ s$^{-1}$ sr$^{-1}$ 
has been derived for 
energies above 100 PeV. 

\begin{figure}[t]
\centering
\includegraphics[width=6cm]{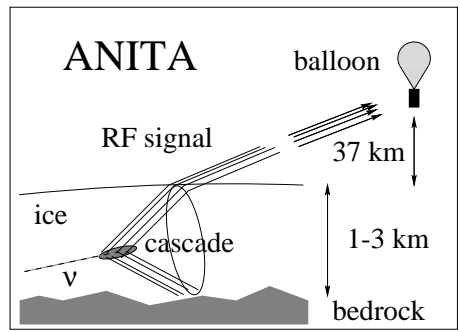}
\vspace{-5mm}
\caption{\label{anita} 
The ANITA balloon project
}
\end{figure}

{\bf SALSA}, a R\&D project study for radio 
detection in natural salt domes, 
promises to get a limit about three orders 
of magnitude better \cite{SALSA}. 
{\bf ANITA}
(ANtarctic Impulsive Transient Array)  
is an array of radio antennas planned to be 
flown at a balloon on an Antarctic circumpolar path in 2006
\cite{ANITA}. 
From 35 km altitude it may 
record the radio pulses from neutrino interactions in the 
thick ice cover and monitor a 
really huge volume (see fig.\ref{anita}). 
The expected sensitivity from a 30 day flight is about 
$10^{-7} E_{\nu}^{-2}$ GeV$^{-1}$ cm$^{-2}$ s$^{-1}$ sr$^{-1}$ 
at 10 EeV.

\begin{figure}
\centering
\includegraphics[width=5.0cm]{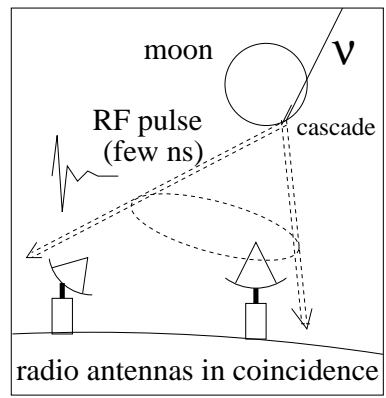}
\vspace{-5mm}
\caption{\label{glue} 
The Goldstone Lunar Ultrahigh Energy neutrino Experiment GLUE.
}
\end{figure}

Most exotic is the Goldstone Lunar Ultrahigh Energy Neutrino 
Experiment, {\bf GLUE} (see fig.\ref{glue}). 
It has searched for radio emission from extremely-high energy 
cascades induced by neutrinos 
or cosmic rays skimming the moon surface \cite{GLUE}. 
Using two NASA antennas, 
an upper limit of 
$10^{-4} E_{\nu}^{-2}$ GeV$^{-1}$ cm$^{-2}$ s$^{-1}$ sr$^{-1}$ 
at 100 EeV has been obtained.

\vspace{-5mm}

\vspace{-5mm}

\section{Detection of neutrino energies via air showers}

\begin{figure}
\centering
\includegraphics[width=7.0cm]{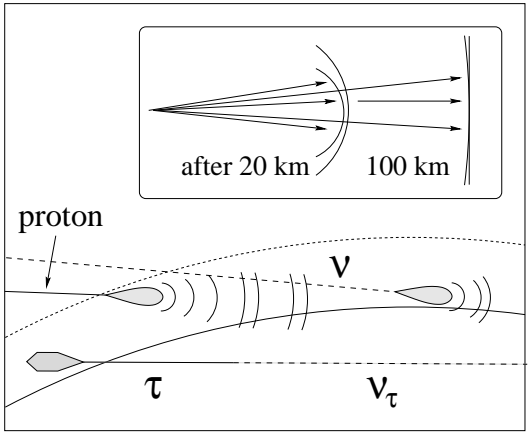}
\vspace{-3mm}
\caption{\label{horizontal} 
Detection of fluorescence light emitted by
horizontal or upward directed air showers from
neutrino interactions.
}
\end{figure}

\begin{figure}
\centering
\includegraphics[width=5.5cm]{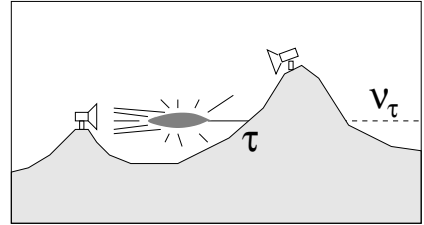}
\vspace{-5mm}
\caption{\label{mount} 
Detection of tau neutrino showers behind a mountain.
}
\end{figure}

At supra-EeV energies, large extensive air shower arrays like the
{\bf AUGER} detector in Argentina \cite{AUGER} or the
telescope array \cite{TA}
may seek for horizontal air showers due to neutrino interactions deep in 
the atmosphere (showers 
induced by charged cosmic rays start on top of the atmosphere).
Figure \ref{horizontal} explains the principle. 
AUGER consists of an array of water tanks going to 
span an area of more than 3000 km$^2$ and 
will record the Cherenkov light of air-shower particles crossing the tanks. 
It is combined with telescopes looking for the atmospheric fluorescence 
light from air showers. The optimum sensitivity window for this 
method is at 1-100 EeV, the effective detector mass is between 1 and 20 Giga-tons, 
and the estimated sensitivity is of the order of  
$10^{-8} E_{\nu}^{-2}$  GeV$^{-1}$ cm$^{-2}$ s$^{-1}$ sr$^{-1}$. 
An even better sensitivity might be obtained for tau neutrinos, $\nu_{\tau}$, 
scratching the Earth and interacting close to the array.
The charged $\tau$ lepton produced in the interaction can escape 
the rock around the array,  in contrast to electrons,
and in contrast to muons it decays after a 
short path into hadrons. 
If this decay happens above the array or in the field of view 
of the fluorescence telescopes, the decay cascade can be recorded. 
Provided the experimental pattern allows clear identification,  
the acceptance for this kind of signals can be large. 
For the optimal energy scale of 1 EeV, the sensitivity might reach  
$10^{-8} E_{\nu}^{-2}$ GeV$^{-1}$ cm$^{-2}$ s$^{-1}$ sr$^{-1}$.
A variation of this idea is to search for tau lepton cascades which are
produced by horizontal PeV neutrinos hitting a  mountain and then
decay in a valley between target mountain and an ``observer'' mountain
-- see fig.\ref{mount} 
\cite{mountain}.

Already eight years ago, the {\bf Fly's Eye} 
collaboration \cite{Fly}, and more recently,
the Japanese {\bf AGASA} collaboration \cite{AGASA}
have practiced the search mode of horizontal 
air showers. AGASA derived an upper limit of the order of 10$^{-5}$ 
in the units given above 
- only just one order of magnitude above some 
predictions for AGN jets and for topological defects.

\begin{figure}
\centering
\includegraphics[width=5.0cm]{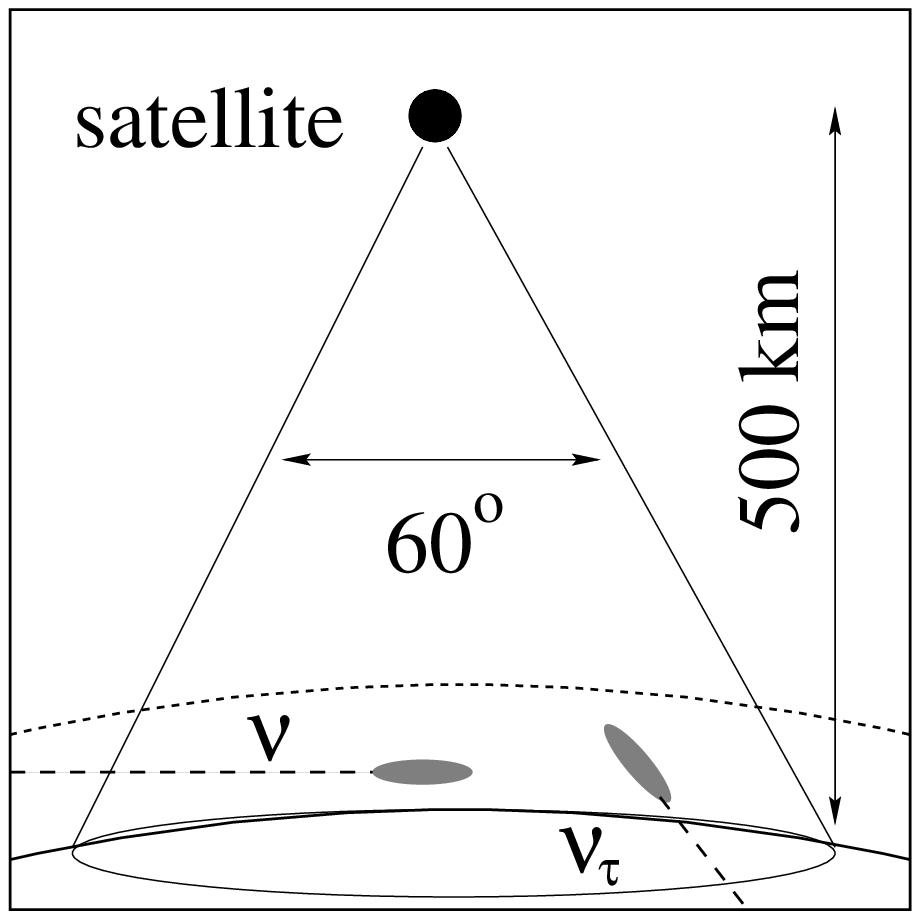}
\vspace{-5mm}
\caption{\label{euso} 
Principle of neutrino detection by satellite
experiments.
}
\end{figure}

Heading to higher energies leads to space based detectors 
monitoring  larger volumes than 
visible from any point on the Earth surface. 
The projects {\bf EUSO} \cite{EUSO} and {\bf OWL} \cite{OWL}  
foresee to launch 
large mirrors with optical detectors to 500 km height. The mirrors 
would look down upon 
the atmosphere and search for nitrogen fluorescence signals 
due to neutrino interactions (see fig.\ref{euso}).  
The monitored mass would be up to  10 Tera-tons, 
with an energy threshold of about 10$^{10}$ GeV.

Finally, I mention the idea to detect the radio emission from
cosmic ray and neutrino induced air showers with low frequency 
radio telescopes, as dicussed in \cite{Lofar}.

\section{Scenario for the next decade}

\begin{figure}[h]
\centering
\vspace{0.8cm}
\includegraphics[width=7.3cm]{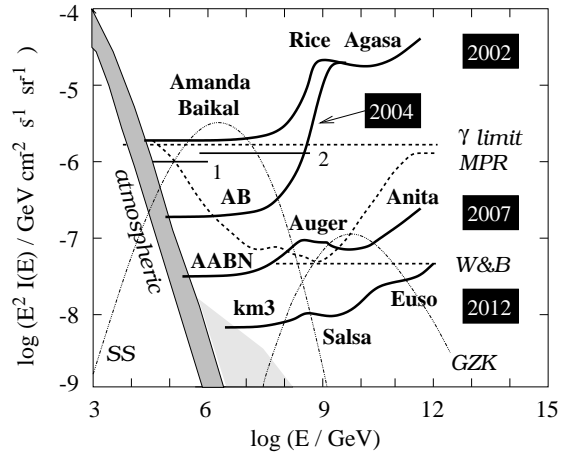}
\vspace{-8mm}
\caption{\label{diffuse} 
Scenario for the improvement of experimental sensitivities 
to  diffuse extraterrestrial fluxes of high energy neutrinos.
AB = Amanda,\,Baikal, AABN = Amanda,\,Antares,\,Baikal,\,Nestor.
{\it 1,2,:} Amanda limits obtained from the analysis of
upward (1) and high energy downward (2) tracks, assuming
an $E^{-2}$ spectrum. 
The grey band denotes the flux of atmospheric neutrinos, with
the excess at high energies being an estimate for the
contribution from prompt muons and neutrinos due to
charm decays in air showers. 
Dashed lines indicate various theoretical bounds, the 2
thin curves specific flux predictions (see text).}
\end{figure}

\begin{figure}[t]
\centering
\includegraphics[width=7.3cm]{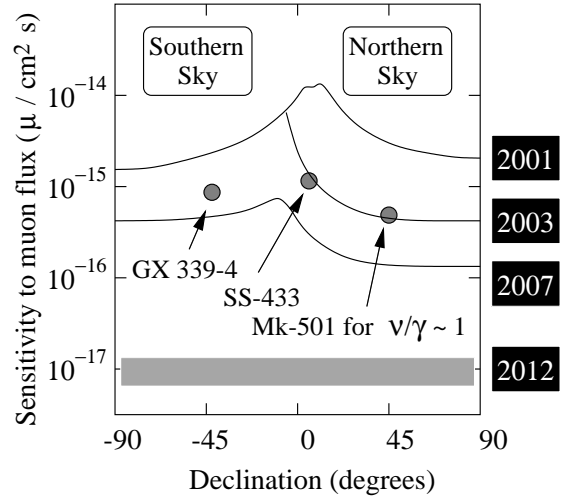}
\vspace{-5mm}
\caption{\label{point} 
Scenario for the 
improvement of experimental sensitivities to TeV point sources. 
Expected steps for the Northern sky are obtained from Amanda (2003), 
and Amanda together with the first strings of IceCube (2007), 
on the Southern sky from the Mediterranean detectors Antares
and Nestor (2007). 
In 2012, both hemispheres will have profited from cubic 
kilometer arrays indicated by the grey band.
Shown are also predicted fluxes for two microquasars \cite{Guetta} - one on
the
northern and one on the southern hemisphere - which are just in
reach for Amanda and the Mediterranean arrays. As a benchmark,
we show also the flux which would be expected if Mk501, a source
spectacular in TeV gamma rays,  would produce a similar
flux in TeV neutrinos.
}
\end{figure}

The next ten years promise to be a particularly 
exciting decade for high energy neutrino astrophysics. 
Figures \ref{diffuse} and \ref{point} sketch possible 
scenarios to move the frontier towards unprecedented  sensitivities. 

Like every estimate of time scales and
expected physics performance, this 
scenario should be taken with caution. 
Notorious time delays in the realization
of projects on the one hand, and possible new approaches
on the other, will likely modify the evolution.
In addition, new techniques have to be fully 
understood -- their sensitivity to signals as well as the
backgrounds!
With AMANDA and BAIKAL, the optical underwater/ice technique
has mastered this phase: downgoing 
muons and atmospheric neutrinos provided an invaluable
calibration source.
The energy range of acoustic or radio techniques as described
in sections 4 and 5, however, is
beyond the range covered by atmospheric neutrinos or 
normal cosmic rays.  Calibration of the detectors in the absence
of a surefire signal will be a challenge.

Figure \ref{diffuse} addresses the sensitivity to diffuse fluxes, 
i.e. integrating over the full angular acceptance of the detectors. 
The scale is set by the known flux of atmospheric neutrinos, 
by the bounds derived from observed fluxes of charged cosmic rays 
(W\&B \cite{WB} and, with less stringent assumptions, 
the lower MPR curve \cite{MPR}), by gamma rays (horizontal
MPR line which assumes that cosmic rays are mostly confined
in the cosmic source region and only gammas and neutrinos escape), 
and by specific model predictions \cite{Semikoz}. The figure shows 
two of the latter, one for the predicted flux of
GZK neutrinos at ultra-high energies (see item {\it b)} above), 
the other for a model on neutrinos from
Active Galactic Nuclei, peaking at TeV-PeV energies (Stecker and
Salomon, SS \cite{SS}).
The majority of the limits shown
are published as ``differential'' limits, defining the
flux sensitivity as the neutrino flux which gives at least one
observed event per decade of energy per year (assuming
negligible background). The limits published for AMANDA
assume an $E^{-2}$ flux. They come from two separate analyses, the
one studying upward muons tracks \cite{upICRC}, the other downward
tracks of very high energy which are unlikely being due
to muons generated in the atmosphere \cite{downICRC}.
Both analyses, however, properly account for the background of these 
atmospheric muons. 
The lines
marked {\it 1} and {\it 2} extend over the energy range containing
90\% of the events expected from an $E^{-2}$ flux.
For better illustration of the
progress in time and over all the energy range,
the AMANDA limits as well as the limits expected for
ICECUBE have been translated to limits differentially
per energy decade.

The present frontier is defined by TeV-PeV 
limits obtained by AMANDA and BAIKAL, and by PeV-EeV limits from the 
South Pole radio array RICE and the Japanese AGASA air shower array. 
Note that the Baikal/Amanda limits just reached a level
sufficient to test (and actually to exclude) the model shown.
The progress over the next 2 years will come from AMANDA and BAIKAL. 
After that, the Mediterranean telescopes -- ANTARES and
NESTOR -- will start to contribute, 
flanked by AUGER and the ANITA balloon mission at high energies. 
This could result in an improvement of about two orders of magnitude 
over the full relevant energy range. 
Actually, five years from now a large variety of models might
have been tested, including several predictions for neutrinos at
GZK energies \cite{Semikoz}.
Finally, in ten years from now, 
the TeV-PeV sensitivity will be defined by the cubic kilometer 
arrays at the South Pole and in the Mediterranean (marked
as ``km3''). 
At the high energy frontier, a SALSA-like experiment, 
and still higher, satellite detectors, might push the limit down by 
about three orders of magnitude compared to 2002.

Most likely, the first signal with clear signature will be a point source, 
possibly a transient signal which is easiest to identify. 
Figure \ref{point} sketches a possible road until 2012. 
Best present limits are from MACRO, Super-Kamiokande (Southern sky) 
and AMANDA-B10 (Northern sky). This picture will not change until 
the Mediterranean detectors come into operation.
AMANDA and ANTARES/NESTOR have the first realistic chance to discover 
an extraterrestrial neutrino source. The ultimate sensitivity for 
the TeV-PeV range is likely reached by the cubic kilometer arrays. 
This scale is set by many model predictions for neutrinos from cosmic 
accelerators or from dark matter decay. However, irrespective of 
any specific model  prediction, these gigantic detectors, 
hundred times larger than AMANDA and thousand times larger than 
underground detectors, will hopefully keep the promise for  
any detector opening  a new window to the Universe: 
to detect {\it unexpected} phenomena.

\vspace{1.0cm}

{\bf Acknowledgment:} I thank S.\,Barwick, P.\,Gorham, M.\,Kowalski
and J.\,Learned for stimulating and helpful discussions. 
Furthermore, I acknowledge useful information and suggestions obtained 
from J.\,Brunner, A.\,Ringwald, D.\,Semikoz, A.\,Tsirigotis and  
Sh.\,Yoshida. This paper is a written version of talks given at the 
18th European Cosmic Ray Conference, Moscow 2002,
and the 8th Topical Seminar on Innovative Particle and Radiation Detectors,
Siena 2002. I thank L.\,Kuzmichev and F.\,Navarria, respectively, 
for their kind invitations and their support.

\end{document}